\documentclass[pra,twocolumn,superscriptaddress,longbibliography]{revtex4-1} 
\usepackage{times, xspace}  
\usepackage{amsbsy,amssymb,amsmath,bm,bbold} 
\usepackage{graphicx,color,epsfig,rotate} 
\usepackage{fancyhdr} 
\usepackage{epstopdf}
\usepackage{float}
\usepackage[colorlinks=true,linkcolor=blue,citecolor=blue,urlcolor=blue]{hyperref}

\usepackage{tikz}

\usepackage[pagewise]{lineno}

\newcommand{\circledT}{%
	\vcenter{\hbox{%
			\tikz[baseline=(c.base)]{
				\node[circle,draw,inner sep=0pt,minimum size=0.46cm,line width=0.5pt] (c)
				{\footnotesize$\mathcal{T}$};
				\draw[line width=0.5pt] (c.north) -- ++(0,0.2);
				\draw[line width=0.5pt] (c.south) -- ++(0,-0.2);
			}
	}}%
}

\def\bbbc{{\mathchoice {\setbox0=\hbox{$\displaystyle\rm C$}\hbox{\hbox 
to0pt{\kern0.4\wd0\vrule height0.9\ht0\hss}\box0}} 
{\setbox0=\hbox{$\textstyle\rm C$}\hbox{\hbox 
to0pt{\kern0.4\wd0\vrule height0.9\ht0\hss}\box0}} 
{\setbox0=\hbox{$\scriptstyle\rm C$}\hbox{\hbox 
to0pt{\kern0.4\wd0\vrule height0.9\ht0\hss}\box0}} 
{\setbox0=\hbox{$\scriptscriptstyle\rm C$}\hbox{\hbox 
to0pt{\kern0.4\wd0\vrule height0.9\ht0\hss}\box0}}}}

\DeclareMathAlphabet\mathbfcal{OMS}{cmsy}{b}{n}

\AtBeginDocument{%
	\newwrite\bibnotes
	\def\bibnotesext{Notes.bib}
	\immediate\openout\bibnotes=\jobname\bibnotesext
	\immediate\write\bibnotes{@CONTROL{REVTEX41Control}}
	\immediate\write\bibnotes{@CONTROL{%
			apsrev41Control,author="08",editor="1",pages="1",title="0",year="1"}}
	\if@filesw
	\immediate\write\@auxout{\string\citation{apsrev41Control}}%
	\fi
}%

\begin{document} 
\title{Dual-state control of lasing and absorption via conjugate exceptional points}
\author{Arnab Laha}
\email{arnablaha777@gmail.com}
\affiliation{Science, Mathematics, and Technology (SMT), Singapore University of Technology and Design (SUTD), Singapore 487372, Singapore}
\author{Somnath Ghosh}
\affiliation{Department of Physics, \'Ecole Centrale School of Engineering, Mahindra University, Hyderabad 500043, India}
\author{Adam Miranowicz}
\affiliation{Institute of Spintronics and Quantum Information, Faculty of Physics and Astronomy, Adam Mickiewicz University, 61-614 Pozna\'n, Poland}
\author{Lin Wu}
\email{lin\_wu@sutd.edu.sg}
\affiliation{Science, Mathematics, and Technology (SMT), Singapore University of Technology and Design (SUTD), Singapore 487372, Singapore}

\date{\today}

\begin{abstract}	
Lasing and coherent perfect absorption (CPA) are time-reversed manifestations of non-Hermitian light-matter interactions. While exceptional points (EPs) have been extensively explored for controlling lasing dynamics, their role in the concurrent manipulation of lasing and absorption remains largely unexplored. Here, we demonstrate the emergence of a pair of conjugate second-order EPs (EP2s) in a gain-loss-engineered Fabry-P\'erot microcavity that enables dual-state operation involving both coherent amplification and absorption. By spatially tailoring gain and loss, we realize two such EP2s: one associated with the coalescence of coupled scattering-matrix poles and the other, its conjugate, arising from the coalescence of corresponding zeros, thereby directly linking the amplifying and absorbing branches of the system. Leveraging the branch-point topology of these conjugate EP2s, we adiabatically encircle them in the gain-loss parameter space and achieve deterministic state permutation, enabling multiple reconfigurable switching schemes for the controlled generation and manipulation of threshold lasing and CPA. Notably, simultaneous encirclement of these conjugate EP2s yields a coordinated dual-state switching protocol, resulting in a frequency-matched coexistence of lasing and absorption responses within the same cavity. Our results establish an EP-based framework for unified and flexible control of lasing and absorption in non-Hermitian photonic systems.
\end{abstract}  
\maketitle %

\section{Introduction}

\begin{figure*}[t]
	\centering
	{\includegraphics[width=12.2 cm]{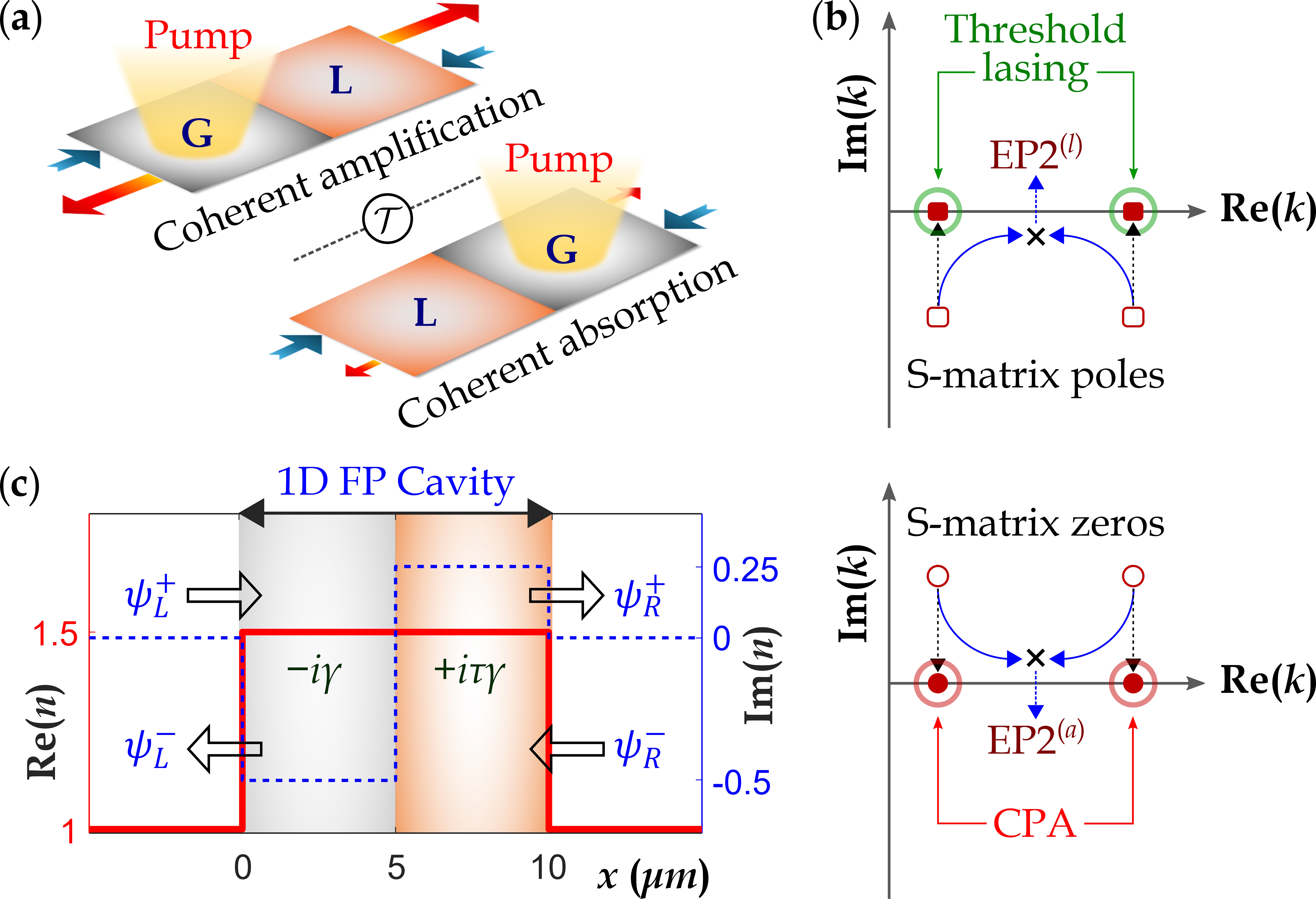}}
	\caption{\textbf{Thematic concept and system design}. 
		\textbf{(a)} Schematic illustration of coherent amplification and absorption as time-reversed ($\mathcal{T}$-symmetric) processes, in which identical incident fields produce amplified or suppressed outputs in the presence of gain (G) or loss (L), respectively. 
		\textbf{(b)} Evolution of $S$-matrix poles and zeros in the complex $k$-plane under different gain-loss configurations. Hollow square and circular markers denote, respectively, a pair of poles (upper panel) and zeros (lower panel) of the passive cavity. Black dotted trajectories indicate their evolution under uniform gain and uniform loss, respectively, as they approach the real $k$ axis, signaling the onset of threshold lasing (upper panel) and CPA (lower panel), marked by the corresponding solid symbols. In contrast, solid blue trajectories illustrate the evolution under tailored (nonuniform) gain-loss distributions, where poles and zeros are driven toward two distinct second-order exceptional points: EP2$^{(l)}$, arising from pole coalescence in the lasing channel, and EP2$^{(a)}$, arising from zero coalescence in the absorbing channel. The superscripts $(l)$ and $(a)$ denote lasing- and absorption-related EPs, respectively.
		\textbf{(c)} Schematic of the gain-loss-engineered one-dimensional Fabry-P\'erot (FP) microcavity, showing the forward- and backward-propagating fields $\psi^{\pm}_{L,R}$ at the cavity interfaces. The corresponding real and imaginary parts of the refractive-index profile $n(x)$ are shown for representative parameters $\gamma=0.5$ and $\tau=0.5$, corresponding to a gain-dominated configuration.}
	\label{fig1}
\end{figure*}

Lasing and its time-reversal ($\mathcal{T}$)-symmetric counterpart, coherent perfect absorption (CPA), constitute two complementary manifestations of non-Hermitian wave physics, reflecting the fundamental duality between optical amplification and absorption \cite{Liertzer12,Wang21CPA,Chong10CPA,Longhi2010_ptla,Chong2011_ptla,Hang_2016}. In open scattering systems, a unified description of coherent amplification and absorption, schematically illustrated in Fig.~\ref{fig1}(a), can be formulated in terms of the evolution of poles and zeros of the associated scattering ($S$) matrix in the complex frequency ($k$) plane, as shown in Fig.~\ref{fig1}(b). While the poles are directly related to the eigenvalue spectrum of an effective non-Hermitian Hamiltonian, a full scattering description naturally incorporates both poles and zeros within a unified framework. These poles and zeros are discrete analytic singularities of the $S$-matrix embedded in a continuous spectrum: poles correspond to quasi-bound resonant states defined by the boundary conditions of purely outgoing fields (amplifying response), whereas zeros correspond to antiresonant scattering conditions defined by the boundary conditions of purely incoming fields (absorbing response) \cite{Binkowski2024_polezero,Fortman2025}. Introducing uniform gain drives the poles toward the real $k$ axis, signaling the onset of threshold lasing associated with self-sustained emission [illustrated by black dotted arrows in Fig.~\ref{fig1}(b)]. Conversely, replacing gain with an equivalent amount of uniform loss pushes the zeros toward the real $k$-axis, giving rise to CPA, where coherently incident waves are completely absorbed \cite{Chong10CPA,Longhi2010_ptla,Chong2011_ptla}. Such poles or zeros residing on the real $k$-axis within the scattering continuum correspond to spectral singularities \cite{Longhi10SS,Mostafazadeh09SS,Hang_2016}.

Within the framework of spectral singularities, the concept of CPA-laser points, most commonly realized using balanced gain and loss in parity-time ($\mathcal{PT}$)-symmetric configurations, has been extensively investigated \cite{Longhi2010_ptla,Chong2011_ptla}. Although CPA-lasers exploit non-Hermitian coupling, their operation is typically limited to a static coincidence of a scattering-matrix pole and zero at a designed operating point. As a result, they lack an intrinsic topological mechanism linking amplification and absorption beyond this isolated condition. This limitation motivates the exploration of exceptional point (EP) singularities, which enable a qualitatively different paradigm for controlling amplification and absorption through nontrivial topology \cite{Midya18,Ganainy18,Miri19,Ozdemir19,Parto21review,Wang23review}. In contrast to spectral singularities, which correspond to isolated real-axis conditions of the scattering spectrum, EPs are branch-point degeneracies in parameter space that govern the global evolution of scattering singularities (poles and zeros) under continuous variation of system parameters. In this context, recent advances on conjugate EPs \cite{Laha22,Laha24} provide a promising route toward topologically coordinated control of lasing and absorption within a unified non-Hermitian framework.

To place this concept in context, we briefly recall the defining features of EP singularities. EPs were originally introduced as defective degeneracies of non-Hermitian operators acting on discrete eigenspaces, where both eigenvalues and their associated eigenvectors coalesce, resulting in a topological defect in the underlying spectrum\cite{Kato,Heiss2004,Minganti2019}. While initially formulated for discrete spectra, the concept of EPs has since been generalized to open resonance systems associated with continuous spectra \cite{Heiss00PRE,Heiss12JPA,Pick17,Eleuch2017,Ge11,Shou2025}. This generalization has found broad realization in photonic platforms, including optical microcavities \cite{Laha20EP3,Laha21EP4,Laha2025,Kullig18njp,Chen17sensor}, waveguides \cite{Laha22,Laha24,Laha18,Zhang18insitu,Zhang19EP3,Laha20}, and photonic crystals \cite{Ding15,Kaminski17,Bykov18}, where EPs manifest through the coalescence of scattering resonances. Beyond their prominent role in ultrasensitive detection \cite{Chen17sensor,Hodaei17sensor,Wiersig20}, the branch-point topology of EPs---revealed through state permutation under adiabatic parameter encirclement \cite{Dembowski04,Gilary13na}---has enabled diverse forms of non-Hermitian light control, including topological state flipping \cite{Laha20EP3,Laha21EP4,Laha2025,Kullig18njp}, chirality-driven mode conversion \cite{Laha18,Zhang18insitu,Zhang19EP3,Doppler16}, and strongly nonreciprocal transmission \cite{Laha20,Choi17isolator}. EPs have also emerged as powerful tools for manipulating lasing \cite{Liertzer12,Kullig2025_lowQ,Ji2023,Benzaouia2022_EPlasing} and CPA \cite{Wang21CPA,Helmut2024_CPA,Suwunnarat2022_EPCPA}, although these two phenomena have largely been investigated separately, with control typically focused on either amplification or absorption.

The spectral topology associated with conjugate EPs \cite{Laha22,Laha24} offers a distinct framework for understanding the interplay between amplification and absorption in non-Hermitian systems. The underlying analytic structure can be described using an effective two-level non-Hermitian Hamiltonian $\mathcal{H}(\lambda)$ governed by a complex control parameter $\lambda=\lambda_{\mathrm{R}}+i\lambda_{\mathrm{I}}$. The corresponding eigenvalues $E_{1,2}(\lambda)$ remain analytic functions of $\lambda$ except at a second-order exceptional point (EP2) located at $\lambda=\lambda^{s}$ in the complex $\lambda$ plane. Here, two complex conjugate conditions can be realized depending on the reversal of $\lambda_{\mathrm{I}}$, giving rise to a pair of conjugate EP2s defined as $\lambda^{s}_{\pm}=\lambda^{s}_{\mathrm{R}}\pm i\lambda^{s}_{\mathrm{I}}$, which are related by $\mathcal{T}$-symmetry \cite{Laha22}. While this Hamiltonian description captures the eigenvalue degeneracies associated with EPs, the corresponding scattering formulation is required to describe both pole and zero branches (i.e., the corresponding spectral solutions of the $S$-matrix) within a single analytic structure. 

When implemented in an open system within the $S$-matrix framework, such conjugate EP2s can arise under spatially nonuniform gain-loss distributions. In this setting, EP2s originate either from the interaction of coupled scattering poles within the amplifying branch or from the interaction of coupled scattering zeros within the absorbing branch of the cavity, as schematically indicated by the solid blue arrows and black crosses in Fig.~\ref{fig1}(b). These conjugate EP2s govern spectral evolution in both gain-dominated and loss-dominated regimes, thereby establishing a topological bridge between the pole and zero branches of the underlying $S$-matrix. As a consequence, this conjugate-EP-mediated pole-zero correspondence provides a natural mechanism for coordinated and reconfigurable control of coherent amplification and absorption within a unified scattering framework. In this context, the present approach introduces a distinct form of dual tunability compared to recent demonstrations of EP-based control of lasing \cite{Kullig2025_lowQ,Ji2023,Benzaouia2022_EPlasing} and CPA \cite{Wang21CPA,Helmut2024_CPA,Suwunnarat2022_EPCPA}, where amplification and absorption are addressed independently.

In this work, we exemplify this concept using a Fabry-P\'erot microcavity with spatially distributed gain and loss, analyzed within the $S$-matrix formalism. Owing to $\mathcal{T}$-symmetry in a two-dimensional gain-loss parameter space, the system exhibits a pair of conjugate EP2s: one associated with the coalescence of scattering poles corresponding to amplifying states, and the other associated with the coalescence of the corresponding zeros describing absorbing scattering conditions. By winding around these conjugate EP2s along suitably designed parametric loops, we reveal coordinated topological evolution of the lasing- and absorption-related states, enabling threshold lasing and CPA under reconfigurable switching, as well as their simultaneous dual-state control. This approach establishes a versatile platform for compact, reconfigurable laser-absorber devices, non-Hermitian logic elements, and advanced nonreciprocal photonic functionalities.

\begin{figure*}[t]
	\centering
	{\includegraphics[width=16.5 cm]{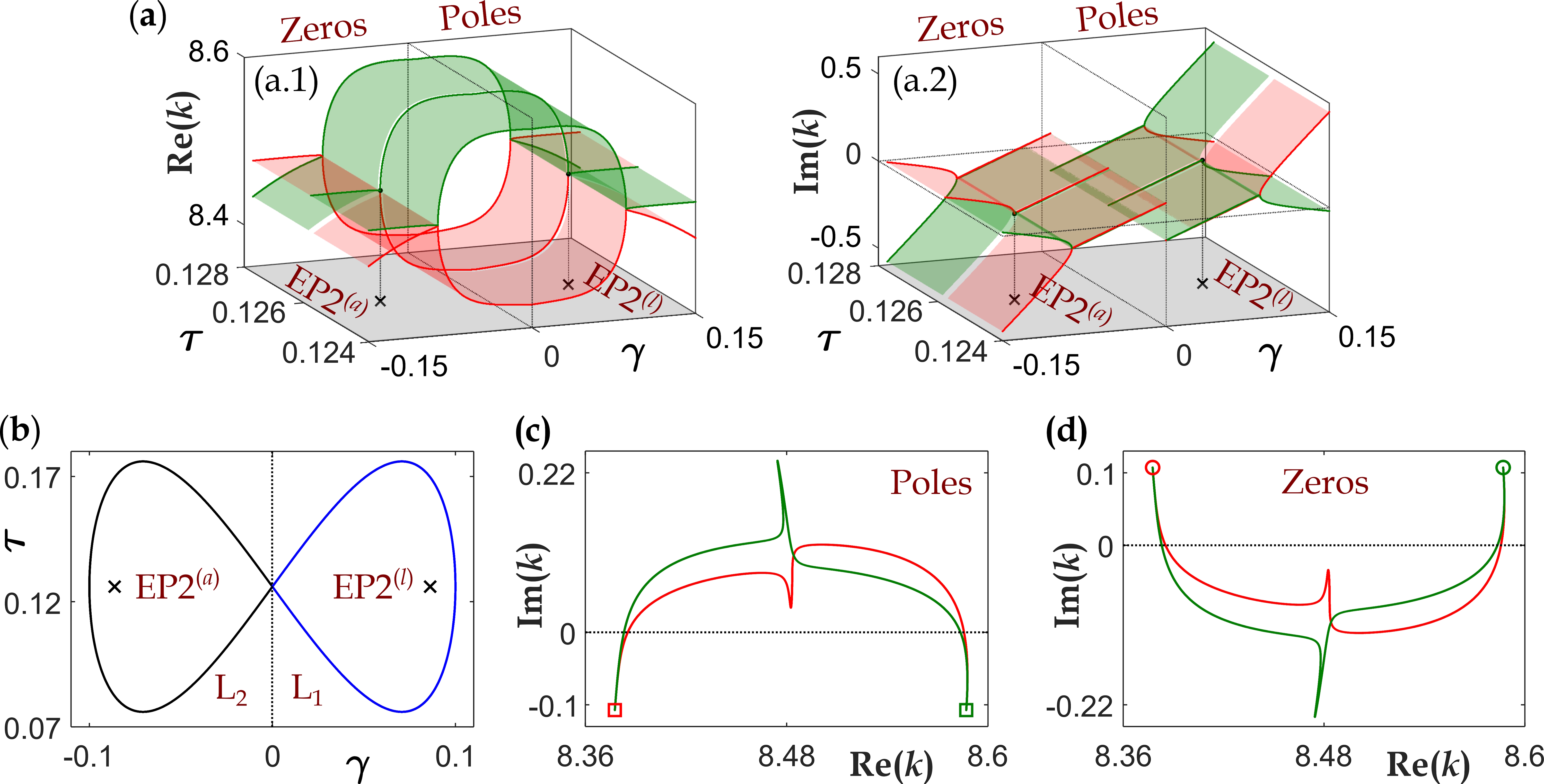}}
	\caption{
		\textbf{Hosting two conjugate EP2s.}
		\textbf{(a)} Riemann surfaces of (a.1) Re($k$) and (a.2) Im($k$), while varying $\gamma$ and $\tau$, showing the evolution of two coupled poles ($\gamma>0$) and two coupled zeros ($\gamma<0$). Red and green curves trace pole/zero trajectories for three representative $\tau$ values, revealing two distinct ARCs at $\tau=0.124$ and $\tau=0.128$, and the emergence of conjugate EP2s (marked by black crosses) at $\tau=0.126$: EP2$^{(l)}$ (pole-coalescence at $\gamma=0.0862$) and EP2$^{(a)}$ (zero-coalescence at $\gamma=-0.0862$). 
		\textbf{(b)} Locations of conjugate EP2s in the $(\gamma,\tau)$-plane (black crosses) and two parametric loops [L$_1$ and L$_2$; following Eq. \eqref{enc1}], where L$_1$ (blue loop) and L$_2$ (black loop) individually encircles EP2$^{(l)}$ and EP2$^{(a)}$, respectively. The dotted line at $\gamma=0$ separates the gain- and loss-dominant regions.
		Trajectories of \textbf{(c)} two coupled poles and \textbf{(d)} two coupled zeros in the complex $k$-plane along the loops L$_1$ and L$_2$, respectively, showing the characteristic topological state permutation features. Red and green colors are used to distinguish between two states in both the pole pair and the zero pair.  In (c) and (d), hollow squares and circles, respectively, represent the initial (passive; $i.e.$, when $\gamma=0$) locations of poles and zeros. Horizontal dotted black lines indicate the real $k$-axis [with $\text{Im}(k)=0$].}
	\label{fig2}
\end{figure*}

\section{Results and Discussion}

\subsection{Gain-loss-engineered microcavity platform}

We consider a one-dimensional (1D) two-port open Fabry-P\'erot microcavity, occupying the region $0\le x\le L$ with $L=10~\mu\mathrm{m}$ and having a uniform passive refractive index, $\mathrm{Re}[n(x)] = 1.5$. This geometry is chosen for its simplicity, compactness, and ease of integration with output coupling along the cavity axis, without requiring phase matching as in ring or toroidal resonators. It also allows clear insight into the interference of longitudinal states distributed along the cavity axis. We introduce non-Hermiticity via an unbalanced bi-layer gain-loss profile (imaginary refractive index), schematically illustrated in Figs.~\ref{fig1}(a) and \ref{fig1}(c), represented by
\begin{equation}
	\text{Im}[n(x)]^{(l|a)}=
	\left\{\hspace*{-0.07cm}\begin{array}{l}-i\gamma\\[3pt]+i\tau\gamma\end{array}\right.
	\hspace*{-0.1cm}\circledT\hspace*{-0.1cm}
	\begin{array}{l}-i(-\gamma)\\[3pt]+i\tau(-\gamma)\end{array}
	\hspace*{-0.07cm}\begin{array}{l}:x\in[0,L/2];\\[3pt]:x\in[L/2,L].
	\end{array}
	\label{nx}
\end{equation}
Here, $\gamma$ and $\tau$ serve as the gain-loss control parameters of the active system. The parameter $\gamma$ sets the overall active strength (negative and positive signs determine the gain and loss, respectively), while $\tau$ controls the gain-loss contrast and thus characterizes the effective openness of the cavity. Depending on the sign of $\gamma$, two distinct active configurations arise, as illustrated in the two panels on the right-hand side of Eq.~\eqref{nx}. In the regime $\tau<1$, $\gamma>0$ corresponds to a gain-dominated cavity (left panel), which supports lasing, whereas $\gamma<0$ corresponds to a loss-dominated cavity (right panel), which supports absorption [consistent with the superscripts $(l|a)$ on the left-hand side of Eq.~\eqref{nx}]. These two configurations are related by $\mathcal{T}$-symmetry. Figure~\ref{fig1}(c) shows the resulting Fabry-P\'erot cavity and its complex refractive-index profile $n(x)$ for representative parameters $\gamma=0.5$ and $\tau=0.5$, corresponding to the gain-dominated cavity conditions.

We employ the $S$-matrix formalism to characterize the spectral response of this non-Hermitian cavity \cite{Ge11,Laha2025}. Denoting $A_L^+$ ($A_R^-$) and $A_L^-$ ($A_R^+$) as the amplitudes of the incident and scattered waves associated with the fields $\psi_L^+$ ($\psi_R^-$) and $\psi_L^-$ ($\psi_R^+$) at the left (right) side of the cavity, respectively, [as indicated in Fig.~\ref{fig1}(c)] the $S$-matrix relation can be written as 
\begin{equation}
	\left[\begin{array}{ll}A_L^-\\[5pt] A_R^+\end{array}\right]
	=S\{n(x),k\}
	\left[\begin{array}{ll}A_L^+\\[5pt] A_R^-\end{array}\right],
	\label{smatrix}
\end{equation}
where, for the present system,
\begin{equation} 
	S(k,\gamma,\tau)=\left[\begin{array}{ll} r_{\mathrm{L}} & t_{\mathrm{LR}} \\[5pt] t_{\mathrm{RL}} & r_{\mathrm{R}} \end{array}\right].
	\label{smatrix}
\end{equation}
The $S$-matrix elements $r_L$ and $r_R$ represent the reflection for left and right incidence, respectively, while $t_{LR}$ and $t_{RL}$ represent the transmission from left to right and from right to left. Within electromagnetic scattering theory, the $S$-matrix can be derived from the transfer matrix $M$, where the overall transfer matrix of a multilayer system is obtained as the product of the transfer matrices of the individual layers. For the designed cavity, the $S$-matrix can be derived from the underlying $2\times2$ transfer matrix $M(k,\gamma,\tau)=[M_{ij}]$ with $i,j\in\{1,2\}$, where the $S$-matrix elements can be expressed as
\begin{equation}
	\begin{array}{ll}
		r_{\mathrm{L}} = -\dfrac{M_{21}}{M_{22}},\qquad 
		& r_{\mathrm{R}}=\dfrac{M_{12}}{M_{22}}, \\[10pt]
		t_{\mathrm{LR}}=\dfrac{1}{M_{22}},\qquad 
		& t_{\mathrm{RL}}=\dfrac{\det(M)}{M_{22}}.
	\end{array}
	\label{s_elements}%
\end{equation}

Within the input--output formalism, the poles of the $S$-matrix are associated with the eigenvalue condition of an effective non-Hermitian Hamiltonian $\mathcal{H}_{\mathrm{eff}}(\gamma,\tau)$ \cite{Chitsazi2014}, satisfying 
\begin{equation}
\det\left[k - \mathcal{H}_{\mathrm{eff}}(\gamma,\tau)\right]=0.
\label{pole}
\end{equation} 
However, the zeros of the $S$-matrix do not arise as inverse solutions of this pole condition; rather, they correspond to distinct spectral conditions and can be related to the corresponding adjoint (time-reversed) Hamiltonian description. Instead, the scattering formulation provides a more general framework that incorporates zeros through the boundary conditions satisfying 
\begin{equation}
\det \left[S(k,\gamma,\tau)\right]=0.
\label{zero}
\end{equation}  
The roots of Eqs.~(\ref{pole}) and (\ref{zero}) define the positions of poles and zeros, respectively, in the complex $k$-plane. Therefore, poles and zeros originate from different spectral conditions and cannot be obtained from a single Hamiltonian eigenvalue problem.

However, within the $S$-matrix formalism, poles and zeros naturally emerge as the fundamental spectral descriptors of the cavity. In practice, these spectral conditions are identified numerically via the divergence and vanishing of the eigenvalues of the $S$-matrix, respectively, leading to:
\begin{subequations}
	\begin{align}
		&\text{Poles:}\qquad\dfrac{1}{\max\big|\text{eig}[S]\big|}=0\\[2pt]
		&\text{Zeros:}\qquad\min\big|\text{eig}[S]\big|=0,
	\end{align}
	\label{pz}%
\end{subequations}
which represent the amplifying and absorbing scattering conditions, respectively, corresponding to two distinct spectral responses of the system. Consistent with energy conservation and causality, the physically acceptable solutions of Eq. \eqref{pz} correspond to the poles located in the fourth quadrant and the zeros (appears as the complex-conjugate counterparts of poles) located in the first quadrant of the complex frequency ($k$)-plane under passive cavity condition ($i.e.$, $\gamma=0$) [as shown in Fig.~\ref{fig1}(b)]. These solutions are approximately linearly spaced according to the longitudinal order $m$, with $\text{Re}(k) \simeq m\pi/[\text{Re}(n)L]$.

\begin{figure*}[t]
	\centering
	{\includegraphics[width=16.5 cm]{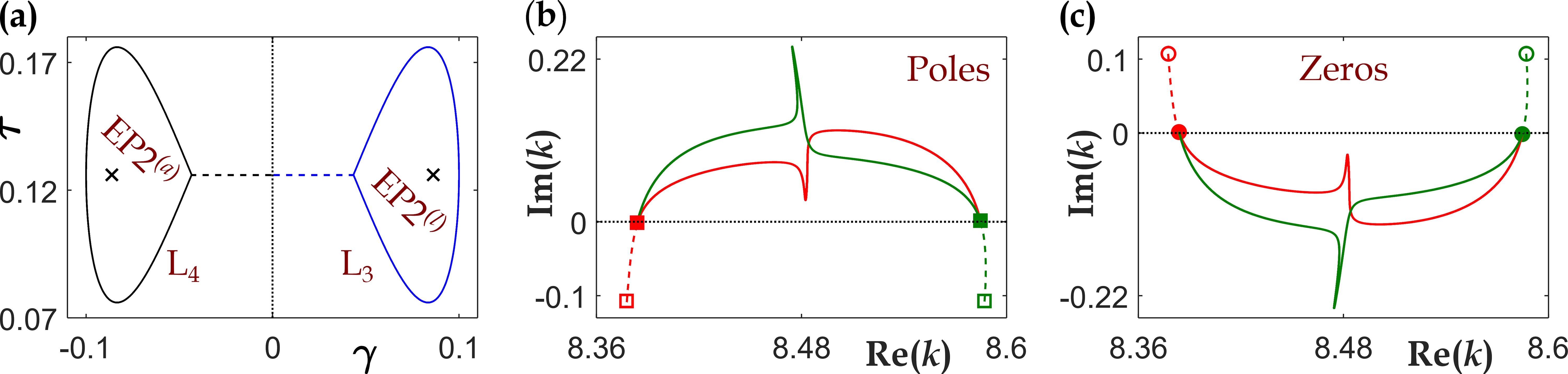}}
	\caption{
		\textbf{Conjugate-EP2-mediated selective control of amplifying and absorbing responses.}
		\textbf{(a)} Individual parametric loops (L$_3$ and L$_4$) in the $(\gamma,\tau)$-plane encircling EP2$^{(l)}$ ($\gamma>0$) and EP2$^{(a)}$ ($\gamma<0$), defined according to Eq.~\eqref{enc2}. In each case, $\gamma$ is first varied linearly to $\pm\gamma_m$ (blue dashed segments) to bring the system to the threshold-lasing or CPA condition before initiating the EP encirclement. 
		\textbf{(b)} Corresponding trajectories of the two coupled poles in the complex $k$-plane along the loop L$_3$ [encircling EP2$^{(l)}$]. Starting from the passive configuration, the poles are first driven to their threshold-lasing conditions [$\mathrm{Im}(k)=0$] via linear variation of $\gamma$ (dashed curves) $\gamma_m$. Subsequent encirclement of EP2$^{(l)}$ induces a topological permutation between those poles (solid curves), enabling controlled conversion between the two threshold lasing states. 
		\textbf{(c)} Corresponding trajectories of the two coupled zeros in the complex $k$-plane along the loop L$_4$ [encircling EP2$^{(a)}$]. The zeros are similarly driven to their respective CPA points via linear variation of $\gamma$ up to $-\gamma_m$ (dashed curves) and subsequent encirclement results in a controlled exchange between the two CPA conditions (solid curves). 
		Hollow square and circular markers indicate the passive ($\gamma=0$) locations of the poles and zeros, respectively, while the corresponding solid markers denote their positions at $\gamma=\pm\gamma_m$ (i.e., their locations on the real $k$ axis). Markers, colors, and symbols not explicitly defined here follow the same conventions as in Fig.~\ref{fig2}.
	}
	\label{fig3}
\end{figure*}

Since the pole and zero branches respond independently to the applied gain and loss, respectively, a patterned gain-loss distribution enables coordinated control of the amplifying and absorbing responses, which forms the central concept of this work. Notably, both branches are embedded within the analytic structure of the $S$-matrix and can be tracked simultaneously within the same parameter space, enabling their coupled evolution in the presence of conjugate EPs. We focus on a specific spectral window $8.36 \le \text{Re}(k) \le 8.6~\mu\mathrm{m}^{-1}$, containing a closely spaced pair of poles and a corresponding pair of zeros. Introducing gain and loss activates mutual coupling within each pair. By systematically varying the gain-loss parameters $\gamma$ and $\tau$, we delve into corresponding interactions by tracking the resulting pole and zero trajectories (governed by the $S$-matrix dependence on $\gamma$ and $\tau$) in the vicinity of two conjugate EP2s.

Although the present analysis focuses on a specific spectral window involving a single pair of poles and zeros, the underlying framework is not restricted to this regime. The microcavity supports multiple poles and zeros of the $S$-matrix, allowing additional conjugate EP pairs to be identified in other spectral regions. More generally, in multimode systems where multiple poles and zeros coexist, richer interaction scenarios can arise, enabling multi-state control associated with more complex topological structures, including multiple coupled EP2s and potentially higher-order EPs. The $S$-matrix-based formulation employed here is applicable to a broad class of non-Hermitian photonic platforms and provides a general basis for exploring such effects. In practice, the spectral window is selected based on the target device operation; the near-infrared regime considered here is relevant for a wide range of photonic systems, and the corresponding pole-zero pairs can be selectively addressed through appropriate tuning of system parameters.

\subsection{Conjugate exceptional points and branch-point topology}

Figure \ref{fig2}(a) presents the Riemann-surface topology of Re($k$) [Fig. \ref{fig2}(a.1)] and Im($k$) [Fig. \ref{fig2}(a.2)], mapped over the 2D $(\gamma,\tau)$ parameter space, showing the evolution of the coupled pole pair and zero pair. Red and green curves distinguish the two individual states in each pair. A vertically oriented dotted rectangle at $\gamma=0$ separates the gain-dominant ($\gamma>0$) and loss-dominant ($\gamma<0$) regimes, where the poles and zeros evolve independently. The avoided resonance crossing (ARC)-type interaction behavior becomes clear along three representative $\tau$ values. At $\tau=0.124$, both the pole-pair and zero-pair exhibit an anticrossing in Re($k$) and a crossing in Im($k$); however, at a slightly higher $\tau=0.128$, this pattern reverses, producing a crossing in Re($k$) and an anticrossing in Im($k$). The transition between these two topologically distinct behaviors occurs at the intermediate value $\tau=0.126$, where the two poles coalesce at $\gamma=0.0862$, and the two zeros coalesce at $\gamma=-0.0862$. These coalescences identify a conjugate pair of EP2s: EP2$^{(l)}$ and EP2$^{(a)}$, located at $(\pm 0.0862, 0.126)$ in the $(\gamma,\tau)$-plane, arising from the inherent pole-zero conjugacy of the scattering system.

To uncover the branch-point topology imposed by the two conjugate EP2s, we track the evolution of poles and zeros under their controlled adiabatic encirclement in the $(\gamma,\tau)$-plane [Figs. \ref{fig2}(b)--\ref{fig2}(d)]. Accordingly, we define two parametrized loops (denoted as L$_1$ and L$_2$)
\begin{subequations}
	\begin{align}
	&\gamma^{\pm}(\phi)=\pm\gamma_0\sin(\phi/2),\\[2pt]
	&\tau(\phi)=\tau_0+a\sin(\phi),
	\end{align}
	\label{enc1}%
\end{subequations}
based on $0\le\phi\le2\pi$. As illustrated in Fig.~\ref{fig2}(b), the positive and negative sign of $\gamma(\phi)$ confine the loop entirely to the gain-dominant ($\gamma>0$) or loss-dominant ($\gamma<0$) regions (visually separated by a black dotted line at $\gamma=0$). This allows L$_1$ and L$_2$ to encircle EP2$^{(l)}$ and EP2$^{(a)}$ individually, facilitating independent evolution of the corresponding poles and zeros. The chosen characteristic parameters, $\gamma_0=0.1$, $\tau_0=0.126$, and $a=0.05$, guarantee that each EP2 lies fully inside its corresponding loop.

\begin{figure*}[t]
	\centering
	{\includegraphics[width=16.5 cm]{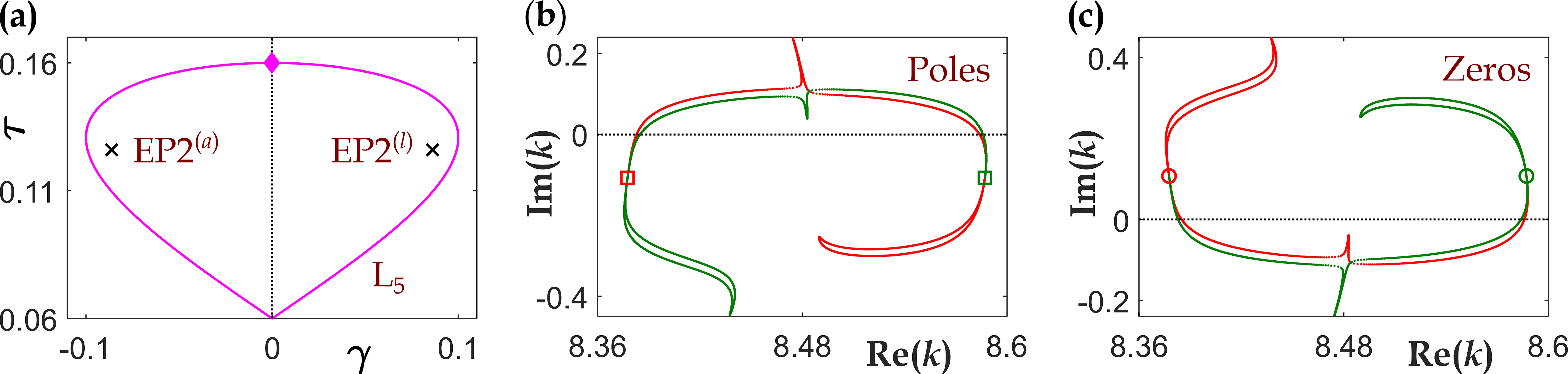}}
	\caption{
		\textbf{Conjugate EP topology for coordinated state control.}
		\textbf{(a)} A single customized parametric loop (L$_5$, shown in magenta) in the $(\gamma,\tau)$-plane, defined by Eq.~\eqref{enc3}, that simultaneously encircles EP2$^{(l)}$ and EP2$^{(a)}$. The magenta diamond marker denotes the passive configuration ($\gamma=0$), which lies on the loop trajectory.
		\textbf{(b)} Corresponding evolution of the two coupled poles in the complex $k$-plane, showing simultaneous EP-induced permutation between two lasing-related states.
		\textbf{(c)} Corresponding evolution of the two coupled zeros in the complex $k$-plane, exhibiting simultaneous permutation between two absorption-related states. Markers, colors, and symbols not explicitly defined here follow the same conventions as in Figs.~\ref{fig2} and \ref{fig3}.}
	\label{fig4}
\end{figure*}

Figures \ref{fig2}(c) and \ref{fig2}(d) show the resulting trajectories of poles and zeros, respectively, in the complex $k$-plane. Encircling EP2$^{(l)}$ leads to a characteristic permutation of the two coupled poles, where they exchange their initial spectral positions after one complete loop. Along this path, each pole intersects the real-$k$ axis (marked by a dotted black line) twice: once before and once after the exchange, resulting in two distinct threshold-lasing points within the spectral window. Analogously, encircling EP2$^{(a)}$ yields the topological exchange for the zero pair, where each zero meets the real-$k$ axis twice, generating distinct CPA conditions visited in alternating order along the loop. These behaviors demonstrate how the conjugate EP2s impose a spectral rearrangement of amplifying and absorbing states under adiabatic parameter cycling. The associated multivalued branch-point topology governs the frequency sequence for threshold-lasing and CPA transitions, while their spectral separation can be precisely tuned based on the loop parameters.

\subsection{Selective topological control of amplification and absorption}

To explore the functional utility of EP topology for selective state control, we investigate whether coherent amplification and absorption can be directly manipulated starting from their respective threshold conditions [Fig.~\ref{fig3}]. To this end, the encircling loops are redefined as
\begin{subequations}
	\begin{align}
	&\gamma^{\pm}(\phi)=\pm\gamma_m+(\gamma_0\mp\gamma_m)\sin(\phi/2),\\[2pt]
	&\tau(\phi)=\tau_0+a\sin(\phi),
	\end{align}
	\label{enc2}%
\end{subequations}
such that the system is first driven to the threshold-lasing or CPA spectral point by linearly tuning $\gamma$ to $\pm\gamma_m$, respectively, after which the EP encirclement is initiated from $\gamma=\pm\gamma_m$. The resulting trajectories in the $(\gamma,\tau)$ parameter plane (denoted as L$_3$ and L$_4$; considering $\gamma_m=0.0434$ together with the previously defined parameters for L$_1$ and L$_2$) are shown in Fig.~\ref{fig3}(a). Notably, these loops remain entirely within the gain- and loss-dominated regions, enabling controlled, state-specific initiation of EP2-driven evolution.

Figure~\ref{fig3}(b) depicts the adiabatic evolution of the two coupled poles in the complex $k$-plane under encirclement of EP2$^{(l)}$ defined by Eq.~\eqref{enc2} (i.e., the loop L$_3$). The poles are first brought to their respective threshold-lasing points and subsequently undergo an EP-induced permutation upon completion of the loop, demonstrating adiabatic switching between the two threshold-lasing states. A symmetric behavior is observed for the corresponding zeros along the parametric loop L$_4$, as shown in Fig.~\ref{fig3}(c): the zeros are first driven to their respective CPA conditions and then permute under encirclement of EP2$^{(a)}$, enabling controlled switching between two CPA points. Together, these symmetric state-switching protocols demonstrate that the proposed scheme supports reconfigurable dual-state operation, allowing selective manipulation of amplifying and absorbing states, even when initiated directly from their respective threshold conditions.

\subsection{Coordinated dual-state control via conjugate exceptional points}

Since the conjugate EP2s exhibit complementary pole and zero dynamics [Figs. \ref{fig2}(b)--\ref{fig2}(d) and \ref{fig3}], coordinated manipulation of lasing and absorption within a single operational cycle is highly desirable for functional photonic applications [Fig.~\ref{fig4}]. To this end, we consider a unified parametric loop (denoted as L$_5$) defined by
\begin{subequations}
	\begin{align}
	&\gamma(\phi)=\gamma_0+a\sin(\phi),\\[2pt]
	&\tau(\phi)=\tau_m+\tau_0\sin(\phi/2),
	\end{align}
	\label{enc3}%
\end{subequations}
with $\gamma_0=0$, $\tau_0=0.1$, $\tau_m=0.06$, and $a=0.1$. As shown in Fig.~\ref{fig4}(a), L$_5$ spans both gain- and loss-dominated regions, thereby simultaneously encircling EP2$^{(l)}$ and EP2$^{(a)}$. This simultaneous encirclement induces a coordinated evolution of the poles and zeros, reflecting their joint embedding within the $S$-matrix structure while arising from distinct spectral conditions. The loop necessarily intersects the passive condition at $\gamma=0$ (marked by the magenta diamond), where it remains mathematically continuous, although this point corresponds to a physical discontinuity under realistic pumping conditions. We emphasize that the EP encirclement process refers to a continuous evolution in parameter space, understood in terms of analytic continuation; since the $\gamma=0$ point on the loop is not a spectral singularity of the $S$-matrix, the resulting spectral behavior depends only on the path taken around the EPs and can be realized, in principle, through segmented or staged parameter variations without modifying the underlying physics.

The resulting evolution of the scattering poles and zeros in the complex $k$-plane under this single-loop perturbation is shown in Figs.~\ref{fig4}(b) and \ref{fig4}(c), respectively. Both poles and zeros undergo EP-induced identity exchange, demonstrating that a single parametric cycle can simultaneously orchestrate the spectral rearrangement of lasing- and absorption-related states. As in the previous cases, each pole and zero intersects the real-$k$ axis twice, giving rise to two distinct threshold-lasing and CPA points. Unlike the individual encirclement schemes discussed earlier, the pole and zero trajectories under this unified loop are nonuniformly distorted and no longer mirror each other.

The origin of this nonuniform and non-mirrored evolution can be understood by examining the underlying parametric dynamics of poles and zeros. The parametric evolution of poles and zeros is governed by the implicit spectral condition
\begin{equation} 
F(k,\gamma,\tau)=0,
\end{equation} 
where $F$ denotes the spectral function corresponding to either poles or zeros (say, $F_p$ and $F_z$, respectively), as defined by Eqs.~(\ref{pole}) and (\ref{zero}). Under parametric variation, the evolution of the corresponding roots follows from implicit differentiation,
\begin{equation}
\nabla_{(\gamma,\tau)}k = -\dfrac{\nabla_{(\gamma,\tau)}F}{\partial_k F}.
\label{rate_eq}
\end{equation}
This relation indicates that the trajectories are governed by how variations in the system parameters modify the underlying gain-loss profile. When the parametric loops are mirror symmetric in the $(\gamma,\tau)$-plane, as shown in Figs.~\ref{fig2}(b) and \ref{fig3}(a), the corresponding refractive index distributions are related by complex conjugation. As a result, the spectral functions associated with poles and zeros exhibit conjugate variation, leading to mirror-symmetric trajectories of the corresponding poles and zeros in the complex $k$-plane with respect to the real axis [Figs.~\ref{fig2}(c, d) and \ref{fig3}(b, c)]. In contrast, when a single loop simultaneously acts on both poles and zeros [Fig.~\ref{fig4}(a)], the gain-loss excursion becomes asymmetric, and the corresponding spectral functions no longer exhibit this conjugate behavior, resulting in non-mirrored trajectories of the corresponding poles and zeros in the complex $k$-plane, as observed in Figs. \ref{fig4}(b) and \ref{fig4}(c).

A notable consequence of this coordinated encirclement is that two pole-zero conjugate pairs---originating from the individual pole and zero pairs---converge to the same resonance frequency, $\mathrm{Re}(k)=8.512~\mu\mathrm{m}^{-1}$, but at opposite gain-loss settings ($\gamma=\pm0.086$, $\tau=0.11$). This behavior implies that, irrespective of whether the system operates in a gain-dominated or loss-dominated regime, it supports the coexistence of lasing- and absorption-associated responses at the same frequency, revealing an unconventional form of dual-state accessibility mediated by conjugate EP2 topology. Such coordinated pole-zero evolution provides a level of control over amplification and absorption that is not accessible within conventional Hamiltonian descriptions (typically associated with a single spectral branch) and goes beyond a simple dual manifestation of a single non-Hermitian degeneracy.

The present analysis is based on a linear scattering description, which is well suited to capture the pole-zero structure, the associated topological features of conjugate EPs, and the corresponding threshold-lasing and CPA conditions. While nonlinear effects, such as gain saturation and nonlinear feedback, usually influence the detailed dynamical behavior of a physical laser system above threshold, \cite{Ramezani2021} the EP structure and its associated topology are governed by the underlying linear operator. Accordingly, the spectral features discussed here are expected to remain valid. Moreover, nonlinear effects can be neglected to first order in the vicinity of threshold-lasing (for poles) or CPA conditions (for zeros). A comprehensive treatment of strongly nonlinear regimes, including effects such as mode competition and dynamical instabilities, is beyond the scope of the present work and may be relevant for detailed device-level investigations.

\section{Conclusion}

In summary, we have demonstrated that a gain-loss-engineered Fabry-P\'erot microcavity can host a pair of conjugate EP2s in its parameter space, providing a unified topological mechanism for the simultaneous and coordinated control of amplification and absorption through the pole-zero correspondence of the underlying $S$-matrix. By tailoring parametric loops within gain- and loss-dominated regions, we realize EP-driven spectral rearrangements of amplifying and absorbing states via independent adiabatic switching of coupled pole and zero pairs. The multivalued branch-point topology associated with the conjugate EP2s naturally gives rise to threshold-lasing and CPA points along these switching pathways. We further demonstrated controlled exchanges between two threshold-lasing states or two CPA states under state-specific loops, and, importantly, achieved coordinated dual-state control using a single-loop operation that encloses both EP2s simultaneously. This approach produces a unique frequency-matched coexistence of lasing-associated and absorption-associated states within the same cavity, revealing an unconventional dual-state accessibility enabled by non-Hermitian topology.

Together, these results establish a versatile and reconfigurable platform for dual-state photonic control, offering a distinct degree of tunability compared to existing EP-based lasers or absorbers, where amplification and absorption are typically addressed independently. Beyond advancing the fundamental understanding of non-Hermitian light-matter interactions, the proposed schemes offer a practical route toward compact photonic devices with integrated emitter-absorber functionalities, logic-driven photonic elements, and dynamically programmable nonreciprocal responses. Looking ahead, the conjugate-EP framework can be extended to higher-dimensional configurations, multimode and coupled-cavity systems, and topologically robust photonic networks, opening new directions for EP-assisted photonics in sensing, quantum information processing, and tunable light-matter interfaces.

\section*{Acknowledgment}
A.L. and L.W. acknowledge the support from National Research Foundation (NRF-CRP26-2021-0004,  NRF-CRP31-0007), Ministry of Education (MOE-MOET32024-0005), Agency for Science, Technology and Research (MTC IRG M24N7c0083), and Singapore University of Technology and Design (SKI 2021-04-12; under Kickstarter Initiative), Singapore.

\bibliography{References}

\end{document}